\documentclass[10pt,twocolumn,letterpaper]{article}
\usepackage[affil-it]{authblk}

\usepackage[pagenumbers]{cvpr} 

\usepackage{graphicx}
\usepackage{amsmath}
\usepackage{amssymb}
\usepackage{booktabs}

\usepackage[pagebackref,breaklinks,colorlinks]{hyperref}

\usepackage[capitalize]{cleveref}
\crefname{section}{Sec.}{Secs.}
\Crefname{section}{Section}{Sections}
\Crefname{table}{Table}{Tables}
\crefname{table}{Tab.}{Tabs.}

\title{Colon Nuclei Instance Segmentation\\ using a Probabilistic Two-Stage Detector}

\author[1, 2]{Pedro Costa}
\author[3]{Yongpan Fu}
\author[1,2]{João D. Nunes}
\author[1,2]{Aurélio Campilho}
\author[1,2]{Jaime S. Cardoso}

\affil[1]{Faculty of Engineering, University of Porto, Porto, Portugal}
\affil[2]{Institute for Systems and Computer Engineering
Technology and Science (INESC TEC), Porto, Portugal}
\affil[3]{Czech Technical University, Faculty of Electrical Engineering, Prague, Czesh Republic}

\date{\today}

\begin{document}

\maketitle

\begin{abstract}
   Cancer is one of the leading causes of death in the developed world. Cancer diagnosis is performed through the microscopic analysis of a sample of suspicious tissue.
   This process is time consuming and error prone, but Deep Learning models could be helpful for pathologists during cancer diagnosis.
   We propose to change the CenterNet2 object detection model to also perform instance segmentation, which we call SegCenterNet2.
   We train SegCenterNet2 in the CoNIC challenge dataset and show that it performs better than Mask R-CNN in the competition metrics. 
\end{abstract}

\section{Introduction}
\label{sec:intro}

Current cancer diagnosis is very time consuming, having very low throughput, and is prone to high inter and intra observer variability \cite{elmore2015diagnostic}.
Parhologists start by performing a biopsy on the suspicious tissue, then the tissue is stained using Hematoxylin and Eosin, and finally the tissue is analyzed under a microscope.

Pathologists look for diverse histological properties of WSIs while searching for signs of cancer. These properties include the organization of cell nuclei, their density and other nuclei morphological features \cite{araujo2017classification}.
These features can then be used to predict survival \cite{alsubaie2018bottom,lu2018nuclear} or characterize the disease \cite{lu2018nuclear}.

Deep Learning models could be used for the detection of individual nuclei and the classification of the type of cell they belong to.
However the development of a Deep Learning model that detects and segments cell nuclei in WSIs is challenging. In particular, we must consider memory constrainst imposed by the high dimensional WSIs. Besides, each image contains a high density of cell nuclei, tipically more than $100$, which is by far superior than the typical number of objects that standard object detection models are trained on.


In this work we implement an instance segmentation model, that simultaneously detects and segments cell nuclei in crops of WSIs.
We propose to use CenterNet2 \cite{zhou2021probabilistic}, a probabilistic two-stage object detection model. 
This model allows the reduction of proposals from the Region Proposal Network (RPN), which could be important in this application where each image has many objects.
We also updated the original model to perform segmentation.
Additionally, we propose a novel Region of Interest (RoI) Head that explores the spatial relationships between objects in the image.






\section{Method}
\subsection{CenterNet2}

The main contribution CenterNet2 was a probabilistic interpretation of two-stage object detection models.
This novel interpretation allowed the proposal of a new training objecive that ties the two-stages together.

The goal of an object detection model is to produce a set of $K$ bounding box detections, with an associated class distribution $P(C_k=c)$ for classes $c$ including the background class.
Two-stage detectors approach this problem by first modeling a class-agnostic object likelihood $P(O_k)$ in the first stage and a conditional categorical classifcation $P(C_k|O_k)$ in the second stage. 
The joint class distribution is as follows:
\begin{equation}
    P(C_k) = \sum_o P(C_k|O_k=o)P(O_k=o),
\end{equation}
where $o$ is $0$ when the object belongs to the background and $1$ when it is a positive detection.

For positive examples, object detection models can be trained by maximizing the following objective:
\begin{equation}
    log P(C_k) = log P(C_k|O_k=1) + log P(O_k=1).
\end{equation}

However, for negative examples, the objective does not factorize, since we may have positive detections in the first stage that are classified as background in the second stage:
\begin{equation}
    log P(bg) = log(P(bg|O_k=1)P(O_k=1) + P(O_k=0)).
\end{equation}

The authors proposed to maximize two lower bounds of $log P(bg)$:

\begin{equation}
    log P(bg) \geq P(O_k=1) log (P(bg|O_k=1)),
\end{equation}
\begin{equation}
    log P(bg) \geq log(P(O_k=0)).
\end{equation}

They jointly optimize these two lower bounds and show improved performance.

The first stage of the detector uses CenterNet \cite{zhou2019objects}, but CenterNet2 uses a ResNet-FPN \cite{lin2017feature} backbone. Then, for the second stage, we use a RoIAlign followed by a standard RoIHead to regress the bounding box location and object class.

\subsection{Instance Segmentation}

CenterNet2 original paper only supports object detection, therefore, we updated CenterNet2's RoI Head to also include mask prediction, following Mask R-CNN's \cite{he2017mask} approach.

For that, we added $3$ additional Fully-Connected layers that map the region of interest feature vector into a fixed size mask of $14 \times 14$px. 
We only predict a single segmentation mask, independent of the object class.
The ground-truth mask is resized to the resolution of the predicted mask before applying an IoU loss.
We call this modified CenterNet2 as SegCenterNet2.

\subsection{Implementation Details}

We used Detectron2 \cite{wu2019detectron2}, a PyTorch \cite{NEURIPS2019_9015} framework to implement and train our models.
We trained the models for 15000 iterations with a batch size of 8.
We used a warm-up learning rate scheduler, where the learning rate is linearly increased during 2000 interations until reaching its final value of $0.02$. The learning rate was divided by $10$ after $12500$ iterations and then again after $14000$.
The Adam optimizer was used with gradient clipping.

The input images were resized to $800\times800$px to increase the spatial resolution of the FPN.
We used random horizontal flips as data augmentation.
At test time, detections with a score higher than $0.5$ were considered.

\section{Evaluation}

\subsection{Dataset}

In this work, we used the Colon Nuclei Identification and Counting Challenge (CoNIC) \cite{graham2021conic} dataset.
The dataset consists of $4981$ non-overlapping Hematoxylin and Eosin stained histology patches obtained in 5 different centers.
Each cell nucleus belongs to one of 6 classes: neutrophil, epithelial, lymphocyte, plasma, eosinophil, and connective.
We used the same split that the authors of the competition provided, making sure that images from the same patient are all in the same set.

\subsection{Instance Segmentation}

To evaluate our model, we use the competition metrics: PQ and multi-PQ. 
In Table \ref{tab:patient_split} we can see that SegCenterNet2 performs slighlty worse than the baseline, however, by only a small margin.


\begin{table}[]
    \centering
    \begin{tabular}{c|cccc}
        Method & PQ  &  Multi-PQ & DQ & SQ\\ \hline
        HoverNet & \textbf{0.6149} & \textbf{0.4998} & - & - \\
        SegCenterNet2 & 0.6067 & 0.4880 & 0.7846 & 0.7400
    \end{tabular}
    \caption{Comparison between our SegCenterNet2 model with HoverNet.}
    \label{tab:patient_split}
\end{table}

{\scriptsize
\bibliographystyle{ieee_fullname}
\bibliography{egbib}
}

\end{document}